\documentclass[11pt,fleqn]{article}

\usepackage{amsmath,amssymb,amsthm,enumerate,cite}

\newcommand{\p}{\partial}
\newcommand{\const}{\mathop{\rm const}\nolimits}

\newcounter{tbn}

\newcounter{mcasenum}

\newtheorem{theorem}{Theorem}

{\theoremstyle{definition}

\newtheorem{note}{Note}
\begin{document}

\par\noindent {\LARGE\bf
Group Analysis of
Variable Coefficient\\ Diffusion--Convection Equations.\\ IV. Potential Symmetries
\par}
{\vspace{4mm}\par\noindent {\bf N.M. Ivanova~$^\dag$, R.O. Popovych~$^\ddag$ and C. Sophocleous~$^\S$
} \par\vspace{2mm}\par}
{\vspace{2mm}\par\noindent {\it
$^\dag{}^\ddag$~Institute of Mathematics of NAS of Ukraine,~3 Tereshchenkivska Str., 01601 Kyiv, Ukraine\\
}}
{\noindent \vspace{2mm}{\it
$\phantom{^\dag{}^\ddag}$~e-mail: ivanova@imath.kiev.ua, rop@imath.kiev.ua
}\par}

{\par\noindent\vspace{2mm} {\it
$^\ddag$~Fakult\"at f\"ur Mathematik, Universit\"at Wien, Nordbergstra{\ss}e 15, A-1090 Wien, Austria
} \par}

{\vspace{2mm}\par\noindent {\it
$^\S$~Department of Mathematics and Statistics, 
University of Cyprus,
CY 1678 Nicosia, Cyprus\\
}}
{\noindent {\it
$\phantom{^\S}$~e-mail: christod@ucy.ac.cy
} \par}

{\vspace{7mm}\par\noindent\hspace*{8mm}\parbox{140mm}{\small
This paper completes investigation of symmetry properties of nonlinear variable coefficient diffusion--convection equations of the form
$f(x)u_t=(g(x)A(u)u_x)_x+h(x)B(u)u_x$ which was started
in~\cite{Ivanova&Popovych&Sophocleous2006Part1,Ivanova&Popovych&Sophocleous2006Part2,Ivanova&Popovych&Sophocleous2006Part3}.
Potential symmetries of equations from the considered class are found and the connection of them with Lie symmetries of diffusion-type equations
is shown. Exact solutions of the Fujita--Storm equation $u_t=(u^{-2}u_x)_x$ are constructed.
}\par\vspace{7mm}}

\section{Introduction}

This paper is the fourth part of a series of works on symmetry properties of nonlinear variable coefficient diffusion--convection equations of the form
\begin{equation} \label{eqDKfgh}
f(x)u_t=(g(x)A(u)u_x)_x+h(x)B(u)u_x,
\end{equation}
where $f=f(x),$ $g=g(x),$ $h=h(x),$ $A=A(u)$ and $B=B(u)$ are arbitrary smooth functions of their variables,
$f(x)g(x)A(u)\!\neq\! 0.$

In the first three parts of the series we studied equivalence transformations of equations~\eqref{eqDKfgh},
performed the complete group classification of class~\eqref{eqDKfgh}, completely classified local and potential conservation laws
of the considered equations, found contractions of their Lie symmetry algebras, solutions,
conservation laws and characteristics of conservation laws.
Using different generalizations of equivalence of conservation laws proposed in~\cite{Ivanova&Popovych&Sophocleous2006Part3},
we constructed a complete list of inequivalent potential systems of equations~\eqref{eqDKfgh}.
Now, in the last part of the series we classify potential symmetries of equations~\eqref{eqDKfgh},
i.e., symmetries of the constructed potential systems that are not projectible on the space of non-potential variables $(t,x,u)$.

The notion of potential symmetries was introduced in~\cite{Bluman&Reid&Kumei1988,Bluman&Kumei1989}.
A system of partial differential equations may admit symmetries of such sort
when some of its equations can be written in a conserved form.
After introducing potentials for equations written in the conserved form as
additional dependent variables, we obtain a new (potential) system of differential equations.
Any local invariance transformation of the obtained system induces
a symmetry of the initial system.
If transformations of some of the ``non-potential" variables
explicitly depend on potentials, this symmetry
is a non-local (potential)
symmetry of the initial system.

Potential symmetries of different subclasses of class~\eqref{eqDKfgh} have been investigated by many authors.
Thus, e.g., in~\cite{Bluman&Kumei1989} potential symmetries of constant coefficient diffusion equations ($f=g=1$, $B=0$) were found,
potential symmetries of equations with $f=g=h=1$ were found in~\cite{Sophocleous1996}.
In~\cite{Popovych&Ivanova2003PETs} potential symmetries of constant coefficient equations~\eqref{eqDKfgh} were classified
with respect to a group of potential equivalence transformations and connection between potential and Lie symmetries was found.
In~\cite{LisleDissertation} a number of results concerning
(local and potential) equivalence transformations of constant coefficient equations~\eqref{eqDKfgh} were obtained.
Potential symmetries of variable coefficient diffusion equations ($B=0$)
were classified in~\cite{Sophocleous2000,Sophocleous2003,Sophocleous2005}.

\begin{note}
Since potential symmetries of linear parabolic equations have been studied in~\cite{Popovych&Kunzinger&Ivanova2007},
in the subsequent analysis we consider only nonlinear equations of form~\eqref{eqDKfgh}.
\end{note}

The paper is organized as follows. First, we briefly recall results on equivalence transformations
and a list of inequivalent potential systems (Section~\ref{SectionOnEquivTransforPotSys}) of equations from class~\eqref{eqDKfgh}.
Preliminary analysis of the derived potential systems in quite a general form is given in Section~\ref{SectionOnPrelimAnalPotSys}.
The complete classification of potential symmetries of equations from class~\eqref{eqDKfgh} is given in Section~\ref{SectionOnPotSymDCEs},
where we also describe a connection between potential and local symmetries of~\eqref{eqDKfgh}.
Although Fujita--Storm equation is a model for a wide range of physically important processes, only few of its exact solutions are known.
Therefore, in Section~\eqref{SectionOnSolutionsOfFujitaEq}, applying potential equivalence transformation,
we adduce some new exact solutions of the Fujita--Storm equation that cannot be obtained by usual Lie symmetry reduction method.

\section{Equivalence transformations and potential systems}\label{SectionOnEquivTransforPotSys}

The usual equivalence group~$G^{\sim}$ of class~\eqref{eqDKfgh} is formed by the nondegenerate point transformations
in the space of~$(t,x,u,f,g,h,A,B)$, which are projectible on the space of~$(t,x,u)$,
i.e. they have the form
\begin{gather}
(\tilde t,\,\tilde x,\,\tilde u)=(T^t,\,T^x,\,T^u)(t,\,x,\,u), \nonumber\\[0.5ex]
(\tilde f,\,\tilde g,\,\tilde h,\,\tilde A,\,\tilde B)=(T^f,\,T^g,\,T^h,\,T^A,\,T^B)(t,\,x,\,u,\,f,\,g,\,h,\,A,\,B),\label{EquivTransformations}
\end{gather}
and transform any equation from class~\eqref{eqDKfgh} for the function $u=u(t,x)$
with the arbitrary elements $(f,g,h,A,B)$
to an equation from the same class for function $\tilde u=\tilde u(\tilde t,\tilde x)$
with the new arbitrary elements~$(\tilde f,\tilde g,\tilde h,\tilde A,\tilde B)$.

\begin{theorem}\cite{Ivanova&Popovych&Sophocleous2004,Ivanova&Popovych&Sophocleous2006Part1}
$G^{\sim}$ consists of the transformations
\begin{gather*}
\tilde t=\delta_1 t+\delta_2,\quad
\tilde x=X(x), \quad
\tilde u=\delta_3 u+\delta_4, \\
\tilde f=\dfrac{\varepsilon_1\delta_1}{X_x} f, \quad
\tilde g=\varepsilon_1\varepsilon_2^{-1}X_x\, g, \quad
\tilde h=\varepsilon_1\varepsilon_3^{-1}h, \quad
\tilde A=\varepsilon_2A, \quad
\tilde B=\varepsilon_3B,
\end{gather*}
where $\delta_j$ $(j=\overline{1,4})$ and $\varepsilon_i$ $(i=\overline{1,3})$ are arbitrary constants,
$\delta_1\delta_3\varepsilon_1\varepsilon_2\varepsilon_3\not=0$, $X$ is an arbitrary smooth function of~$x$, $X_x\not=0$.
\end{theorem}

Class~\eqref{eqDKfgh} admits other equivalence transformations which do not belong to~$G^{\sim}$
and form, together with usual equivalence transformations, an {\it extended equivalence group}.
We demand for these transformations to be point with respect to $(t,x,u)$.
The explicit form of the new arbitrary elements~$(\tilde f,\tilde g,\tilde h,\tilde A,\tilde B)$ is determined
via $(t,x,u,f,g,h,A,B)$ in some non-fixed (possibly, nonlocal) way.

\begin{theorem}\cite{Ivanova&Popovych&Sophocleous2004,Ivanova&Popovych&Sophocleous2006Part1}
The complete extended equivalence group~$\hat G^{\sim}$ is formed by the transformations
\begin{gather*}
\tilde t=\delta_1 t+\delta_2,\quad
\tilde x=X(x), \quad
\tilde u=\delta_3 u+\delta_4, \\
\tilde f=\dfrac{\varepsilon_1\delta_1\varphi}{X_x}f, \quad
\tilde g=\varepsilon_1\varepsilon_2^{-1}X_x\varphi\,g, \quad
\tilde h=\varepsilon_1\varepsilon_3^{-1}\varphi\,h, \quad
\tilde A=\varepsilon_2 A, \quad
\tilde B=\varepsilon_3 (B+\varepsilon_4 A),
\end{gather*}
where $\delta_j$ $(j=\overline{1,4})$ and $\varepsilon_i$ $(i=\overline{1,4})$ are arbitrary constants,
$\delta_1\delta_3\varepsilon_1\varepsilon_2\varepsilon_3\not=0$,
$X$ is an arbitrary smooth function of~$x$, $X_x\not=0$,
$\varphi=e^{-\varepsilon_4\int \frac{h(x)}{g(x)}dx}$.
\end{theorem}

Group~$\hat G^{\sim}$ contains a (normal) subgroup of gauge equivalence transformations
\begin{gather}
\tilde f=\varepsilon_1\varphi\, f, \quad
\tilde g=\varepsilon_1\varepsilon_2^{-1}\varphi\, g, \quad
\tilde h=\varepsilon_1\varepsilon_3^{-1}\varphi\, h,\quad
\tilde A=\varepsilon_2 A, \quad
\tilde B=\varepsilon_3 (B+\varepsilon_4 A),
\label{GaugeEquivTransformationsDKfgh}
\end{gather}
where $\varphi=e^{-\varepsilon_4\int \frac{h(x)}{g(x)}dx}$,
$\varepsilon_i$ $(i=\overline{1,4})$ are arbitrary constants, $\varepsilon_1\varepsilon_2\varepsilon_3\not=0$
(the variables $t$, $x$ and $u$ do not transform!).

The transformations~\eqref{GaugeEquivTransformationsDKfgh} act only on arbitrary elements
and do not really change equations.
Application of ``gauge'' equivalence transformations is equivalent to rewriting equations
in another form.

The factor-group $\hat G^{\sim}/\hat G^{\sim g}$ coincides for class~\eqref{eqDKfgh} with~$G^{\sim}/G^{\sim g}$
and can be assumed to consist of the transformations
\begin{equation} \label{RealEquivTransformationsDKfgh}\arraycolsep=0em
\begin{array}{l}
\tilde t=\delta_1 t+\delta_2,\quad
\tilde x=X(x), \quad
\tilde u=\delta_3 u+\delta_4,\\[1ex]
\tilde f=\dfrac{\delta_1}{X_x} f, \quad
\tilde g=X_x\,g, \quad
\tilde h=h, \quad
\tilde A=A, \quad
\tilde B=B,
\end{array}
\end{equation}
where $\delta_i$ ($i=\overline{1,4}$) are arbitrary constants, $\delta_1\delta_3\not=0$,
$X$ is an arbitrary smooth function of~$x$, $X_x\not=0$.

Using the transformation $\tilde t=t$, $\tilde x=\int \frac{dx}{g(x)}$, $\tilde u=u$
from $G^{\sim}/G^{\sim g}$, we can reduce equation~(\ref{eqDKfgh}) to
\[
\tilde f(\tilde x)\tilde u_{\tilde t}= (A(\tilde u)
\tilde u_{\tilde x})_{\tilde x} + \tilde h(\tilde x)B(\tilde u)\tilde u_{\tilde x},
\]
where $\tilde f(\tilde x)=g(x)f(x)$, $\tilde g(\tilde x)=1$ and $\tilde h(\tilde x)=h(x)$.
(Likewise any equation of form~\eqref{eqDKfgh} can be reduced to the same form with $\tilde f(\tilde x)=1.$)
That is why, without loss of generality, we can restrict ourselves to the investigation of the equation
\begin{equation} \label{eqDKfh}
f(x)u_t=\left(A(u)u_x \right)_x + h(x)B(u)u_x.
\end{equation}

Any transformation from~$\hat G^{\sim}$, which preserves the condition $g = 1$, has the form
\begin{equation} \label{EquivTransformationsDKfh}\arraycolsep=0em
\begin{array}{l}
\tilde t=\delta_1 t+\delta_2,\quad
\tilde x=\delta_5 \int e^{\delta_8\int\! h}dx+\delta_6, \quad
\tilde u=\delta_3 u+\delta_4,\\[1ex]
\tilde f=\delta_1\delta_5^{-1}\delta_9 fe^{-2\delta_8\int\! h}, \quad
\tilde h=\delta_9\delta_7^{-1} he^{-\delta_8\int\! h}, \\[1ex]
\tilde A=\delta_5\delta_9A, \quad
\tilde B=\delta_7(B+\delta_8A),
\end{array}
\end{equation}
where $\delta_i$ ($i=\overline{1,9}$) are arbitrary constants, $\delta_1\delta_3\delta_5\delta_7\delta_9\not=0$.
(Here and below $\int\! h=\int\! h(x)\,dx$.)

\begin{note}
If $B=0$, we assume $h=1$.
\end{note}

\begin{note}
It has been shown in~\cite{Popovych&Ivanova2004ConsLawsLanl,Ivanova&Popovych&Sophocleous2006Part3} that the equivalence group for a class of
systems or the symmetry group for single system can be prolonged to potential variables for any step of
the direct iteration procedure. It is natural to use the prolonged equivalence groups to classify
possible conservation laws and potential systems in each iteration.
Additional equivalences which exist in some subclasses of the class or arise
after introducing potential variables can be used for further analysis of connections between conservation laws.
\end{note}


In the third part~\cite{Ivanova&Popovych&Sophocleous2006Part3} of the given series we found all possible
inequivalent (with respect to~$\hat G^{\sim}$ and symmetry transformations of the corresponding classes of equations)
simplest potential systems of nonlinear equations~\eqref{eqDKfh}:
\\[1ex]
\noindent{\bf 1.} $h=1$:  \qquad $v_x=fu$, $v_t=Au_x+\int B$
\\[1ex]
{\bf 2.} $A=1$, $B_u\ne0$, $f=-h(h^{-1})_{xx}$:  \qquad
$v_x=e^t(h^{-1})_{xx} u$,\quad $v_t=e^t(-h^{-1}u_x+(h^{-1})_xu-\int\!\! B).$
\\[1ex]
{\bf 3.} $B=1$, $f=h_x$: \qquad $v_x=e^th_xu$, $v_t=e^t(Au_x+hu).$
\\[1ex]
{\bf 4.} $B=1$, $f=h_x+hx^{-1}$: \qquad $v_x=e^txfu$, $v_t=e^t(xAu_x+xhu-\int\!\! A).$
\\[1ex]
\noindent{\bf 5.}  $B=0$: \qquad $v_x=xfu$,\quad $v_t=xAu_x-\int A$
\\[1ex]
\noindent{\bf 6.1.} $B=1$, $f=e^{-\mu/x}x^{-3}$, $h=e^{-\mu/x}x^{-1}$, $\mu\in\{0,1\}$:
\begin{gather*}\textstyle
v_x=e^{\mu t}xfu, \quad v_t=e^{\mu t}x(Au_x+hu)-e^{\mu t}\int\!\! A,
\end{gather*}
\noindent{\bf 6.2.} $B=1$, $f=e^{-\mu/x}x^{-3}$, $h=e^{-\mu/x}x^{-1}$, $\mu\in\{0,1\}$:
\begin{gather*} \textstyle
v_x=e^{\mu t}(tx-1)fu , \quad v_t=e^{\mu t}(tx-1)(Au_x+hu)-te^{\mu t}\int\!\! A
\end{gather*}
\noindent{\bf 7.1.} $B=1$, $f=|x-1|^{\mu-3/2}|x+1|^{-\mu-3/2}$, $h=|x-1|^{\mu-1/2}|x+1|^{-\mu-1/2}$:
\begin{gather*}\textstyle
v_x=e^{2\mu t}(x\cosh t-\sinh t)fu,\quad v_t=e^{2\mu t}(x\cosh t-\sinh t)(Au_x+hu)-e^{2\mu t}\cosh t\int\!\! A.
\end{gather*}
\noindent{\bf 7.2.} $B=1$, $f=|x-1|^{\mu-3/2}|x+1|^{-\mu-3/2}$, $h=|x-1|^{\mu-1/2}|x+1|^{-\mu-1/2}$:
\begin{gather*}\textstyle
v_x=e^{2\mu t}(x\sinh t-\cosh t)fu ,\quad v_t=e^{2\mu t}(x\sinh t-\cosh t)(Au_x+hu)-e^{2\mu t}\sinh t\int\!\! A.
\end{gather*}
\noindent{\bf 7.3.} $B=1$, $f=|x-1|^{\mu-3/2}|x+1|^{-\mu-3/2}$, $h=|x-1|^{\mu-1/2}|x+1|^{-\mu-1/2}$:
\begin{gather*}\textstyle
v_x=e^{(2\mu+1)t}(x-1)fu,\quad v_t=e^{(2\mu+1)t}(x-1)(Au_x+hu)-e^{(2\mu+1)t}\int\!\! A,
\end{gather*}
\noindent{\bf 8.} $B=1$, $f=e^{\mu\arctan x}(x^2+1)^{-3/2}$, $ h=e^{\mu\arctan x}(x^2+1)^{-1/2}$:
\begin{gather*}\textstyle
v_x=e^{\mu t}(x\cos t+\sin t)fu ,\quad v_t=e^{\mu t}(x\cos t+\sin t)(Au_x+hu)-e^{\mu t}\cos t\int\!\! A.
\end{gather*}

For some subclasses of~\eqref{eqDKfh} the spaces of local conservation laws are two-dimensional,
that allowed us to construct general potential systems by means of introducing two potentials for each case simultaneously:

\noindent{\bf 5$'$.}  $B=0$: \qquad  $v_x=fu$, $v_t=Au_x$, $w_x=xfu$, $w_t=xAu_x-\int\!\! A$.
\\[1ex]
{\bf 6$'$.} $B=1$, $f=e^{-\mu/x}x^{-3}$, $h=e^{-\mu/x}x^{-1}$, $\mu\in\{0,1\}$:
\begin{gather*}\textstyle
v_x=e^{\mu t}xfu, \quad v_t=e^{\mu t}x(Au_x+hu)-e^{\mu t}\int\!\! A
\\ \textstyle
w_x=e^{\mu t}(tx-1)fu , \quad w_t=e^{\mu t}(tx-1)(Au_x+hu)-te^{\mu t}\int\!\! A.
\end{gather*}

\noindent{\bf 7$'$.} $B=1$, $f=|x-1|^{\mu-3/2}|x+1|^{-\mu-3/2}$, $h=|x-1|^{\mu-1/2}|x+1|^{-\mu-1/2}$:
\begin{gather*}\textstyle
v_x=e^{(2\mu+1)t}(x-1)fu,\quad v_t=e^{(2\mu+1)t}(x-1)(Au_x+hu)-e^{(2\mu+1)t}\int\!\! A
\\  \textstyle
w_x=e^{(2\mu-1)t}(x+1)fu ,\quad w_t=e^{(2\mu-1)t}(x+1)(Au_x+hu)-e^{(2\mu-1)t}\int\!\! A.
\end{gather*}

\noindent{\bf 8$'$.} $B=1$, $f=e^{\mu\arctan x}(x^2+1)^{-3/2}$, $ h=e^{\mu\arctan x}(x^2+1)^{-1/2}$:
\begin{gather*}\textstyle
v_x=e^{\mu t}(x\cos t+\sin t)fu ,\quad v_t=e^{\mu t}(x\cos t+\sin t)(Au_x+hu)-e^{\mu t}\cos t\int\!\! A,
\\ \textstyle
w_x=e^{\mu t}(x\sin t-\cos t)fu, \quad w_t=e^{\mu t}(x\sin t-\cos t)(Au_x+hu)-e^{\mu t}\sin t\int\!\! A.
\end{gather*}

For more details on equivalence relations on the set of potential systems
and rigorous definitions and proofs see~\cite{Ivanova&Popovych&Sophocleous2006Part3}.

In the subsequent analysis we classify potential symmetries of equations~\eqref{eqDKfh}
arising from the above potential symmetries with respect to (trivial) prolongation of transformations from~$\hat G^{\sim}$
to the corresponding potential variables.

\section{Preliminary analysis of potential systems}\label{SectionOnPrelimAnalPotSys}

As one can see, all simplest potential systems of equations~\eqref{eqDKfh} have the similar structure:
\begin{equation}\label{sysSimpPotSysGenForm}
v_x=\alpha(t,x)u,\quad v_t=\beta(t,x)A(u)u_x+\gamma(t,x,u),
\end{equation}
where $\alpha\beta\ne0$. Here $\alpha(t,x)$ is nothing but the characteristic of the corresponding conservation law.

The initial equation on $u$ is a differential consequence of system~\eqref{sysSimpPotSysGenForm}.
Another differential consequence is the equation
\begin{equation}\label{EqGenSimplestPotEq}
v_t=\beta\left(\frac{v_x}\alpha\right)_x+\gamma\left(t,x,\frac{v_x}\alpha\right)
\end{equation}
on the potential dependent variable~$v$ which is called the potential equation associated with the equation~\eqref{eqDKfh}
and the characteristic~$\alpha$.

Consider a Lie symmetry operator $Q=\tau\p_t+\xi\p_x+\eta\p_u+\theta\p_v$ of system~\eqref{sysSimpPotSysGenForm}.
The coefficients of~$Q$ are functions of $t$, $x$, $u$ and $v$.
The infinitesimal invariance criterion implies for system~\eqref{sysSimpPotSysGenForm} the following system of determining equations
\begin{gather}
\tau_u=\xi_u=\theta_u=0,\label{EqTuple1OfDetEqsForLieSymmetriesOfGenSimplestPotSystem}\\[.5ex]
\tau_x=\tau_v=0,\label{EqTuple2OfDetEqsForLieSymmetriesOfGenSimplestPotSystem}\\[.5ex]
\eta=\frac1\alpha\Bigl(-\xi_v\alpha^2u^2-(\alpha_t\tau+\alpha\xi_x+\alpha_x\xi-\alpha\theta_v)u+\theta_x\Bigr),
\label{EqTuple3OfDetEqsForLieSymmetriesOfGenSimplestPotSystem}\\[.5ex]
\eta\frac{A_u}A= -\eta_u+\theta_v+\xi_x-\frac{\xi\beta_x}\beta-\tau_t-\frac{\tau\beta_t}\beta,
\label{EqTuple4OfDetEqsForLieSymmetriesOfGenSimplestPotSystem}\\[.5ex]
-\beta A(\eta_x+\alpha\eta_v u)+\gamma(\theta_v-\tau_t-\alpha\xi_vu)-\gamma_u\eta-\gamma_x\xi-\gamma_t\tau-\alpha\xi_tu+\theta_t=0.
\label{EqTuple5OfDetEqsForLieSymmetriesOfGenSimplestPotSystem}
\end{gather}

The subsystem~\eqref{EqTuple1OfDetEqsForLieSymmetriesOfGenSimplestPotSystem} of determining equations means that
any Lie symmetry transformation of~\eqref{sysSimpPotSysGenForm} with respect to $t$, $x$ and $v$ does not depend on~$u$.
Equation~\eqref{EqGenSimplestPotEq} is a differential consequence of system~\eqref{sysSimpPotSysGenForm},
and there exists one-to-one correspondence between the sets of solutions of equation~\eqref{EqGenSimplestPotEq} and system~\eqref{sysSimpPotSysGenForm}.
Therefore, the truncated operator $\hat Q=\tau\p_t+\xi\p_x+\theta\p_v$ is a Lie symmetry operator of equation~\eqref{EqGenSimplestPotEq}.

And vice versa, consider a Lie symmetry operator $\hat Q=\tau\p_t+\xi\p_x+\theta\p_v$ of equation~\eqref{EqGenSimplestPotEq}.
The coefficients of~$\hat Q$ are functions of $t$, $x$ and $v$.
Then the prolonged to~$u$ operator $Q=\hat Q+\eta\p_u$, where $\eta$ is defined
by formula~\eqref{EqTuple3OfDetEqsForLieSymmetriesOfGenSimplestPotSystem}, is a Lie symmetry operator of system~\eqref{sysSimpPotSysGenForm}.

%

General potential systems of equation~\eqref{eqDKfh} look like
\begin{gather}
\nonumber
v_x=\alpha(t,x)u,\quad v_t=\beta(t,x)A(u)u_x+\gamma(t,x,u),\\
w_x=\lambda(t,x)u,\quad w_t=\mu(t,x)A(u)u_x+\nu(t,x,u),
\label{sysGenPotSysGenForm}
\end{gather}
where $\alpha\beta\lambda\mu\ne0$, $\alpha$ and $\lambda$ are linearly independent.

Applying Lie infinitesimal criterion for system~\eqref{sysGenPotSysGenForm} being invariant with respect
to the symmetry generator $Q=\tau\p_t+\xi\p_x+\eta\p_u+\theta\p_v+\zeta\p_w$,
similarly to the case of the simplest potential system, we get
\begin{gather*}
\tau_u=\xi_u=\theta_u=\zeta_u=0,\\
\tau_x=\tau_v=\tau_w=0.
\end{gather*}

All potential systems of forms~\eqref{sysSimpPotSysGenForm} are~\eqref{sysGenPotSysGenForm} are constructed with usage of local
conservation laws of equations~\eqref{eqDKfh} which were classified with respect to~$\hat G^{\sim}_1$.
Each of these subclasses of conservation laws is invariant with respect to some subgroup of~$\hat G^{\sim}_1$.
In each case this subgroup has a very simple structure and can be singled out from~\eqref{EquivTransformationsDKfh}
imposing additional condition $\delta_7=\delta_8=0$. It becomes the usual equivalence group of class~\eqref{eqDKfh}
and is trivially prolonged to potentials. Henceforth we will call such prolongation as $G^{\sim}_{\rm pr}$.

More detailed analysis of determining equations for coefficients of symmetry generators for both simplest and general potential
systems requires direct substitution of arbitrary elements $\alpha$, $\beta$, $\gamma$, $\mu$, $\nu$ and~$\psi$.
We skip the intermediate cumbersome calculations and adduce immediately the final results.
Namely, in the next section we list all possible $G^{\sim}_{\rm pr}$-inequivalent {\it pure} potential symmetries of equations~\eqref{eqDKfh},
i.e., symmetries of the derived potential systems that are not projectible on the space of non-potential variables $(t,x,u)$.

\section{Potential symmetries and connections with local symmetries}\label{SectionOnPotSymDCEs}

Now we will investigate Lie symmetries of each of these potential systems separately.
Below we analyze these symmetries adducing only the cases when the Lie symmetries of the systems induce nonlocal symmetries of the initial equation.
At first, let us state that potential systems {\bf 2}, {\bf 6.1}--{\bf 7.3}, {\bf6$'$}--{\bf8$'$}
do not yield any potential symmetry for equations~\eqref{eqDKfh}.

\medskip

\noindent{\bf Potential system~1} associated to the simplest characteristic~$1$ arises in case $h=1$ and,
as we pointed out above, has the form
\begin{equation}\label{SystemPotSysChar1h1}\textstyle
v_x=fu,\quad v_t=Au_x+\int B.
\end{equation}
Its constant coefficient case ($f=1$) has been studied in~\cite{Sophocleous1996,Popovych&Ivanova2003PETs}.
In particular, ibid we performed the complete group classification of system~\eqref{SystemPotSysChar1h1}$|_{f=1}$ and
proved that all the potential symmetries can be obtained
from Lie symmetries of equations~\eqref{eqDKfh}$|_{f=1}$ by means of prolongation to the potential $v$
and application of potential equivalence transformation
\[
\tilde t=t,\quad
\tilde x=x+\varepsilon v,\quad
\tilde u=\dfrac u{1+\varepsilon u},\quad
\tilde v=v,
\]
and potential hodograph transformation
\begin{gather}\label{pothodograph}
\tilde t=t,\quad
\tilde x=v,\quad
\tilde u=u^{-1},\quad
\tilde v=x.
\end{gather}
That is why for the constant coefficient case of system~\eqref{SystemPotSysChar1h1}  below we adduce only the list of systems
and their Lie symmetries which induce potential symmetries of equations from class~\eqref{eqDKfh} and do not consider their connection
with Lie symmetries of diffusion equations.
Moreover, we assume that connection of a potential symmetry of a variable coefficient equation
to Lie symmetry is found by reducing the corresponding potential system to a system~\eqref{SystemPotSysChar1h1}$|_{f=1}$.

The list of~$G^{\sim}_{\rm pr}$-inequivalent constant coefficient systems~\eqref{SystemPotSysChar1h1} inducing potential symmetries
of equations from class~\eqref{eqDKfh} consists of the following ones:

\noindent1.\quad $f=1$, $A=u^{-2}e^{1/u}$, $B=0$:\quad
$A^{\max}=\langle \p_t,\; \p_x,\; \p_v,\; 2t\p_t+x\p_x+v\p_v,\; t\p_t-v\p_x+u^2\p_u\rangle$;

\medskip

\noindent2.\quad $f=1$, $A=\dfrac {u^\mu}{(u+1)^{\mu+2}}$, $B=0$:\\
\phantom{2.\quad}  $A^{\max}=\langle \p_t,\; \p_x,\; \p_v,\; 2t\p_t+x\p_x+v\p_v,\; \mu t\p_t+v\p_x-u(u+1)\p_u-v\p_v\rangle$;

\medskip

\noindent3.\quad $f=1$, $A=\dfrac{e^{\mu\arctan u}}{u^2+1}$,  $B=0$:\\
\phantom{3.\quad} $A^{\max}=\langle \p_t,\; \p_x,\; \p_v,\;2t\p_t+x\p_x+v\p_v,\; \mu t\p_t+v\p_x-(u^2+1)\p_u-x\p_v\rangle$;

\medskip

\noindent4.\quad $f=1$, $A=u^{-2}$,  $B=0$:\\
\phantom{4.\quad} $A^{\max}=\langle \p_t,\;$ $\p_v,\;$ $2t\p_t+u\p_u+v\p_v,\;-vx\p_x+u(ux+v)\p_u+2t\p_v,\;$\\
\phantom{4.\quad} $4t^2\p_t-(v^2+2t)x\p_x+u(v^2+6t+2xuv)\p_u+4tv\p_v,\; x\p_x-u\p_u,\;$ $\phi\p_x-\phi_vu^2\p_u\rangle$;

\medskip

\noindent5.\quad $f=1$, $A=u^{-2}e^{\mu/u}$, $B=(1/u-1)e^{1/u}$:\\
\phantom{5.\quad}
$A^{\max}=\langle \p_t,\; \p_x,\; \p_v,\; (\mu-2)t\p_t+((\mu-1)x+v)\p_x-u^2\p_u+(\mu-1)v\p_v\rangle$;

\medskip

\noindent6.\quad $f=1$, $A=u^{-2}e^{1/u}$, $B=-u^{-2}$:\\
\phantom{6.\quad} $A^{\max}=\langle \p_t,\; \p_x,\; \p_v,\; t\p_t+(x+v)\p_x-u^2\p_u+(v-2t)\p_v\rangle$;

\medskip

\noindent7.\quad $f=1$, $A=\dfrac {u^\mu}{(u+1)^{\mu+2}}$,  $B=-\dfrac {\nu u^{\nu-1}}{(u+1)^{\nu}}$:\\
\phantom{6.\quad} $A^{\max}=\langle \p_t,\; \p_x,\; \p_v,\; (\mu-2\nu)t\p_t+((\mu-\nu)x-v)\p_x+u(u+1)\p_u+(\mu-\nu+1)v\p_v\rangle$;

\medskip

\noindent8.\quad $f=1$, $A=\dfrac {u^\mu}{(u+1)^{\mu+2}}$  $B=-(u+1)\ln\dfrac{u}{u+1}-\dfrac1{u+1}$:\\
\phantom{8.\quad} $A^{\max}=\langle \p_t,\; \p_x,\; \p_v,$ $\mu t\p_t+(\mu x+v-t)\p_x+u(u+1)\p_u+(\mu+1)v\p_v\rangle$;

\medskip

\noindent9.\quad $f=1$, $A=\dfrac{e^{\mu\arctan u}}{u^2+1}$, $B=-\dfrac{(x+\nu)e^{\nu\arctan u}}{\sqrt{u^2+1}}\,$:\\
\phantom{9.\quad}
$A^{\max}=\langle \p_t,\; \p_x,\; \p_v,\;
(\mu-2\nu)t\p_t+((\mu-\nu)x-v)\p_x+(u^2+1)\p_u+(x+(\mu-\nu)v)\p_v\rangle$;

\medskip

\noindent10.\quad $f=1$, $A=u^{-2}$, $B=-u^{-2}$:\\
\phantom{10.\quad} $A^{\max}=\langle \p_t,\; \p_v,\; 2t\p_t+u\p_u+v\p_v,\; -v\p_x+u^2\p_u+2t\p_v,$\\
\phantom{10.\quad} $4t^2\p_t-(v^2+2t)\p_x+2u(uv+2t)\p_u+4tv\p_v,\; \p_x,\; e^{-x}\phi\p_x+e^{-x}(\phi-u\phi_v)u\p_u\rangle$;

\medskip

\noindent11.\quad $f=1$, $A=1$,  $B=2u$:\\
\phantom{11.\quad} $A^{\max}=\langle \p_t,\; \p_x,\; 2t\p_t+x\p_x-u\p_u,\; 2t\p_x-\p_u-x\p_v,$\\
\phantom{11.\quad} $4t^2\p_t+4tx\p_x-2(x+2ut)\p_u-(x^2+2t)\p_v,\; \p_v,\; e^{-v}(h_x-hu)\p_u+e^{-v}h\p_v\rangle$.

\medskip
Here $\mu,\nu=\const$. $(\mu,\nu)\not=(-2,-2),\,(0,1)$ and $\nu\not=-1,0$
for case~7.
$\mu\not=-2,0$ for case~2.
The functions $\phi=\phi(t,v)$ and $h=h(t,x)$ are arbitrary solutions
of the linear heat equation ($\phi_t=\phi_{vv};$ $h_t=h_{xx}$).

Up to the group~$G^{\sim}_{\rm pr}$ there exists exactly one system of form~\eqref{SystemPotSysChar1h1} with non-constant value of~$f$
yielding potential symmetries of equation~\eqref{eqDKfh}, namely,
\begin{equation}\label{SystemPotSysChar1Fx-43Au-2B0}
v_x=x^{-4/3}u,\quad v_t=u^{-2}u_x
\end{equation}
associated to the diffusion equation $x^{-4/3}u_t=(u^{-2}u_x)_x$.
The corresponding algebra of potential symmetries is
\[
\langle\p_t,\,\p_v,\, 3x\p_x-u\p_u-2v\p_v,\, 2t\p_t+u\p_u+v\p_v,\,3xv\p_x-(v+3x^{-1/3}u)u\p_u-v^2\p_v\rangle.
\]
Application of potential hodograph transformation maps system~\eqref{SystemPotSysChar1Fx-43Au-2B0}
to the system
\[
\tilde u=\tilde v^{-4/3}\tilde v_{\tilde x},\quad \tilde v_{\tilde t}=(\tilde v^{-4/3}\tilde v_{\tilde x})_{\tilde x}.
\]
Therefore, we establish the nonlocal mapping between the equations
$x^{-4/3}u_t=(u^{-2}u_x)_x$ and $\tilde v_{\tilde t}=(\tilde v^{-4/3}\tilde v_{\tilde x})_{\tilde x}$ belonging to class~\eqref{eqDKfh}.
(In implicit form this transformation was written firstly in~\cite{Munier&Burgan&Gutierres&Fijalkow&Feix1981}.)
The same transformation establishes isomorphism between the above algebra of the potential symmetries
and maximal Lie invariance algebra
\[
\langle \p_t,\;\p_x,\;2t\p_t+x\p_x,\;4t\p_t+3v\p_v,\;x^2\p_x-3xv\p_v \rangle
\]
of the equation $\tilde v_{\tilde t}=(\tilde v^{-4/3}\tilde v_{\tilde x})_{\tilde x}$.
Tildes over variables in the Lie symmetry operators are omitted.

Considering this together with results of~\cite{Sophocleous1996,Popovych&Ivanova2003PETs} we can state that
potential symmetries of equations~\eqref{eqDKfh} obtained from system~\eqref{SystemPotSysChar1h1} are completely classified
and connection between the potential symmetries and Lie symmetries of equations~\eqref{eqDKfh} are found.

\medskip

\noindent{\bf Potential symmetries obtained from system~3}.
It is possible to introduce potential system associated with the characteristic~$e^t$ of form
\[
v_x=e^th_xu,\quad v_t=e^t(Au_x+hu).
\]
for subclass $B=1$, $f=h_x$ of class~\eqref{eqDKfh}.
There exist five $\hat G^{\sim}_{\rm pr}$-inequivalent values
of arbitrary elements for which system~{\bf 3} gives potential symmetries of the initial equations.
The first four of them are quite similar:

\noindent1.\quad $f=1$, $h=x$, $A=u^{-2}e^{-1/u}$:\\
\phantom{1.\quad} $A^{\max}=\langle e^{-2t}(\p_t-x\p_x),\,e^{-t}\p_x,\,\p_v,\,\p_t+v\p_v,\,(x-2e^{-t}v)\p_x+2u^2\p_u+v\p_v\rangle$;

\medskip

\noindent2.\quad $f=1$, $h=x$, $A=\frac{u^\mu}{(u+1)^{\mu+2}}$:\\
\phantom{2.\quad} $A^{\max}=\langle e^{-2t}(\p_t-x\p_x),\,e^{-t}\p_x,\,\p_v,\,\p_t+v\p_v,\,
-\p_t+(\mu x-2e^{-t}v)\p_x+2u(u+1)\p_u+(\mu+1)v\p_v\rangle$;

\medskip

\noindent3.\quad $f=1$, $h=x$, $A=\frac{e^{\mu\arctan u}}{u^2+1}$:\\
\phantom{3.\quad} $A^{\max}=\langle e^{-2t}(\p_t-x\p_x),\,e^{-t}\p_x,\,\p_v,\,\p_t+v\p_v,\,
(\mu x-2e^{-t}v)\p_x+2(u^2+1)\p_u-(\mu v-2e^tx)\p_v\rangle$;

\medskip

\noindent4.\quad $f=1$, $h=x$, $A=u^{-2}$:\\
\phantom{4.\quad} $A^{\max}=\langle e^{-2t}(\p_t-x\p_x),\,\p_v,\,x\p_x-u\p_u,\,\p_t+v\p_v,\,
-vx\p_x+(v+e^txu)u\p_u+e^{2t}\p_v,$\\
\phantom{4.\quad} $e^{2t}\p_t-(2e^{2t}+v^2)x\p_x+(3e^{2t}+2e^txuv+v^2)u\p_u+2e^{2t}v\p_v,\,e^{-t}\varphi\p_x-\varphi_vu^2\p_u\rangle$.
\\
Here $\varphi=\varphi(t,v)$ is an arbitrary solution of equation $e^{-2t}\varphi_t=\varphi_{vv}$.

\smallskip

Using the local transformation $\tilde t=e^{2t}/2$, $\tilde x=xe^t$, $\tilde u=u$ we can reduce them, and moreover, any equation of form
$u_t=(A(u)u_x)_x+xu_x$
to the constant coefficient equation
$
\tilde u_{\tilde t}=(\tilde A(u)\tilde u_{\tilde x})_{\tilde x}.
$
The same transformation trivially prolonged to the potential $\tilde v=v$ maps the corresponding potential system
\[
v_x=e^tu,\quad v_t=e^t(Au_x+hu)
\]
to the constant coefficient potential system
\[
\tilde v_{\tilde x}=\tilde u,\quad \tilde v_{\tilde t}=A(\tilde u)\tilde u_{\tilde x}
\]
associated to the characteristic~1 (potential system {\bf 1}) that have been studied in~\cite{Bluman&Kumei1989,Sophocleous1996,Popovych&Ivanova2003PETs}.

The last nontrivial example of the systems~{\bf3} giving  potential symmetries of equations~\eqref{eqDKfh} is

\medskip

\noindent5.\quad $f=h_x$, $h=x^{-1/3}$, $A=u^{-2}$:\\
\phantom{5.\quad} $A^{\max}=\langle\p_t+v\p_v,\,\p_v,\,2t\p_t+3x\p_x-u\p_u,\,e^{2t}(\p_t+3x\p_x),\,
3xv\p_x+u(e^tx^{-1/3}u-v)\p_u-v^2\p_v\rangle$.
\\
These values of arbitrary elements correspond to the equation
\[
-x^{-4/3}u_t/3=(u^{-2}u_x)_x+x^{-1/3}u_x.
\]
Using the point transformation $\tilde t=3(e^{-2t}-1)/2$, $\tilde x=xe^{-3t}$, $\tilde u=u$ one can reduce it to equation
\[
\tilde x^{-4/3}\tilde u_{\tilde t}=(\tilde u^{-2}\tilde u_{\tilde x})_{\tilde x}
\]
The same transformation trivially prolonged to the potential variable $\tilde v=v$
maps the corresponding potential system to~\eqref{SystemPotSysChar1Fx-43Au-2B0}, considered earlier.

\vspace{0.5cm}

\noindent{\bf Potential symmetries obtained from system~4}
for subclass $B=1$, $f=h_x+hx^{-1}$ of class~\eqref{eqDKfh} are ones of the most interesting
in the sense of reducibility of them to the local symmetries.
Namely there exist two inequivalent systems of form~{\bf 4} giving potential symmetries of equations~\eqref{eqDKfh}:

\noindent1.\quad $h=x^{1/3}$, $A=u^{-2}$:\\
\phantom{1.\quad }$A^{\max}=\langle\p_t+v\p_v,\,\p_v,\, 3x\p_x-2u\p_u+2v\p_v,\, e^{-t}(4\p_t-3x\p_x),\,
3xv\p_x-2u(v+2e^tx^{4/3}u)\p_u+v^2\p_v\rangle$.
\\
The point transformation $\tilde t=(e^{3t/2}-1)/2$, $\tilde x=xe^{3t/4}$, $\tilde u=u$ maps this equation to
\[
\tilde x^{-2/3}\tilde u_{\tilde t}=(\tilde u^{-2}\tilde u_{\tilde x})_{\tilde x}
\]
and prolonged transformation $\tilde v=v$ maps the corresponding potential system to
system {\bf5}.1 ($c_1=0$, $c_2=(4/3)^{-3/4}$) that will be considered below.

\medskip

\noindent2.\quad $h=\ln|x|/x$, $A=u^{-2}$:\quad $A^{\max}=\langle\p_t+v\p_v,\,\p_v,\, e^{-t}x\p_x,\,
e^{-t}xv\p_x-u^2\p_u+2e^t\p_v\rangle$

Transformation $\tilde t=t$, $\tilde x=v$, $\tilde u=1/u$, $\tilde v=x$ maps the given potential system
to the 
\[
\frac{\tilde v_{\tilde x}}{\tilde v}=e^{-\tilde t}\tilde u,\quad
\frac{\tilde v_{\tilde t}}{\tilde v}=e^{\tilde t}\tilde u_{\tilde x}-\ln|\tilde v|-\tilde u^2,
\]
that is a potential system of generalized Burgers equation
$e^{-\tilde t}\tilde u_{\tilde t}=e^{\tilde t}\tilde u_{\tilde x\tilde x}-2\tilde u\tilde u_{\tilde x}$.
Transforming additionally time variable as $\hat t=e^{\tilde t}$ we obtain
generalized Burgers equation
\[
\tilde u_{\hat t}=\hat t\tilde u_{\tilde x\tilde x}-2\tilde u\tilde u_{\tilde x}
\]
studied at~\cite{Kingston&Sophocleous1991} with three-dimensional symmetry algebra
$
\langle\p_{\tilde x},\,\hat t\p_{\hat t}+\tilde x\p_{\tilde x},\,\hat t\p_{\tilde x}+\frac12\p_{\tilde u} \rangle.
$
Note, that the projection of the operator~$\p_{\tilde v}$ on the space of variables $(\tilde t, \tilde x, \tilde u)$ is trivial.

\medskip

\noindent{\bf Potential symmetries obtained from system~5.}
Consider the last simplest potential system
\begin{equation}\label{SystemPotSysCharxB0}
\textstyle
v_x=xfu,\quad v_t=xAu_x-\int A
\end{equation}
associated to the characteristic~$x$ corresponding to the subclass  $B=0$ of equations~\eqref{eqDKfh}.
The following two cases correspond to the nontrivial potential symmetries of the initial equation.

\noindent1.\quad $f=(c_1x^{3/2}+c_2x^{1/2})^{-4/3}$, $A=u^{-2}$:\quad $A^{\max}=$\\
\phantom{1.\quad} $\langle\p_t,\,\p_v,\,4c_2t\p_t-3(c_1x+c_2)x\p_x+(4c_2+3c_1x)u\p_u,\, 3(c_1x+c_2)x\p_x-(2c_2+3c_1x)u\p_u+2c_2v\p_v,$
\phantom{1.\quad} $3(c_1x+c_2)xv\p_x-(3x^{4/3}(c_1x+c_2)^{-1/3}u+(2c_2+3c_1x)v)u\p_u+c_2v^2\p_v \rangle$

Usual potential hodograph transformation~\eqref{pothodograph} maps arbitrary system of form~\eqref{SystemPotSysCharxB0}
to
\[
\tilde v_{\tilde x}=\frac{\tilde u}{\tilde vf},\quad \tilde v_{\tilde t}=\tilde v\tilde u_{\tilde x}-\tilde u\tilde v_{\tilde x}.
\]
Eliminating variable $\tilde u$ we get a constant coefficient equation
$-(1/\tilde v)_{\tilde t}=(f(\tilde v)\tilde v_{\tilde x})_{\tilde x}$. After changing the variable $\hat v=1/\tilde v$ we obtain
constant coefficient equation
\[
\hat v_{\tilde t}=(\hat f(\hat v)\hat v_{\tilde x})_{\tilde x}
\]
having the form~\eqref{eqDKfh}. Here $\hat f=\hat v^{-2}f(1/\hat v)$.
Application of this transformation to system with $f=(c_1x^{3/2}+c_2x^{1/2})^{-4/3}$
leads to $\hat f=(c_2\hat v+c_1)^{-4/3}$.

\medskip

\noindent2.\quad $f=x^{-2}$, $A=u^{-2}$:\quad
 $A^{\max}=\langle \p_t,\,\p_v,\,x\p_x,\,2t\p_t+u\p_u+v\p_v,\, xv\p_x-u^2\p_u+2t\p_v,$\\
\phantom{3.\quad} $4t^2\p_t+(v^2+2t)x\p_x+(4t-2uv)u\p_u+4tv\p_v,\,x^2\varphi\p_x-xu(\varphi+\varphi_vu)\p_u\rangle$.\\
Here $\varphi=\varphi(t,v)$ is an arbitrary solution of equation $\varphi_t=\varphi_{vv}$.

Local transformation $\tilde t=t$, $\tilde x=-1/x$, $\tilde u=xu$, $\tilde v=v$ maps the equation
$x^{-2}u_t=(u^{-2}u_x)_x$ into $\tilde u_{\tilde t}=(\tilde u^{-2}\tilde u_{\tilde x})_{\tilde x}$ and the given potential
system to
\[
\tilde v_{\tilde x}=\tilde u,\quad \tilde v_{\tilde t}=\tilde u^{-2}\tilde u_{\tilde x},
\]
which is potential system of the constant coefficient equation associated to the characteristic~1.


\medskip

\noindent{\bf Potential symmetries obtained from system 5$'$: $B=0$:}
\begin{equation}\label{SystemGenPotSysChar1x}\textstyle
 v_x=fu,\quad v_t=Au_x,\quad w_x=xfu,\quad w_t=xAu_x-\int\!\! A.
\end{equation}

Under the transformation $\tilde v=v$, $\tilde w=xv-w$ this system is locally equivalent to the second-level
potential system
\begin{equation}\label{SystemPotSysSecondLevel}
\textstyle
\tilde v_x=fu,\quad \tilde w_x=\tilde v,\quad \tilde w_t=\int\!\! A  
\end{equation}
that can be obtained from the conservation law $D_t(v)-D_x(\int\!\! A)=0$ of the potential system~\eqref{SystemPotSysChar1h1}.

Below we adduce classification result for system~\eqref{SystemPotSysSecondLevel} omitting the tildes over the variables

\noindent1.\quad $f=x^{-6}$, $A=u^{-2/3}$: \\
\phantom{1.\quad} $\langle\p_t,\,\p_{v}+x\p_w,\,\p_{w},\,2t\p_t-3x\p_x-15u\p_u-3w\p_{w},\,
4t\p_t-3x\p_x-12u\p_u+3v\p_{v},$\\
\phantom{1.\quad}$x^2\p_x+3xu\p_u+(w-xv)\p_v+xw\p_w,\,xw\p_x-3(xv-2w)u\p_u-v(xv-w)\p_{v}+w^2\p_{w}\rangle$;

\medskip

\noindent2.\quad $f=1$, $A=u^{-2/3}$:\\
\phantom{2.\quad} $\langle\p_t,\; \p_x,\; \p_v+x\p_w,\; \p_w,\; 2t\p_t+x\p_x+v\p_v+2w\p_w,\;
2t\p_t+3u\p_u+3v\p_v+3w\p_w,$\\
\phantom{2.\quad} $w\p_x-3uv\p_u-v^2\p_v\rangle$;

\medskip

\noindent3.\quad $f=x^{-2}$, $A=u^{-2}$: \\
\phantom{3.\quad} $\langle\p_t,\,\p_{w},\,x\p_x-v\p_v,\,2t\p_t+x\p_x+u\p_u+w\p_w,$\\
\phantom{3.\quad} $x(2xv-w)\p_x-u(xv+2u)\p_u+v(w-xv)\p_v+(x^2v^2-2t)\p_w,$\\
\phantom{3.\quad} $4t^2\p_t+x(6t+3x^2v^2+-4xvw+w^2)\p_x+2u(2t-3xuv+2uw-x^2v^2+xvw)\p_u$\\
\phantom{3.\quad} $+(2xvw-2t-x^2v^2-w^2)\p_v+2(2tw+x^3v^3-x^2v^2w)\p_w,$\\
\phantom{3.\quad} $x^2\rho_z\p_x+(\rho_z-u\rho_{zz})xu\p_u+\rho\p_v+x(\rho+xv\rho_z)\p_w\rangle$.\\
Here $z=w-xv$, $\rho=\rho(t,z)$ is an arbitrary solution of the linear heat equation $\rho_t=\rho_{zz}$.

\medskip


\noindent4.\quad $f=1$, $A=u^{-2}$:\\[1ex]
\phantom{4.\quad} $\langle\p_t,\,\p_{v}+x\p_w,\,2t\p_t+x\p_x+v\p_{v}+2w\p_{w},\, x\p_x-u\p_u+w\p_w$,\\[1ex]
\phantom{4.\quad} $(w-2vx)\p_x+(2xu^2+uv)\p_u+2t\p_v+(2tx-xv^2)\p_w,\,\lambda_v\p_x-\lambda_tu^2\p_u+(v\lambda_v-\lambda)\p_w$,\\[1ex]
\phantom{4.\quad} $2t^2\p_t+(vw-\frac{3}2xv^2-3tx)\p_x+(u^2(3xv-w)+\frac{uv^2}2+5tu)\p_u+2tv\p_v+(\frac{1}2v^2w-tw-xv^3)\p_w\rangle$\\[1ex]
Here $\lambda=\lambda(t,v)$ is an arbitrary solution of the equation $\lambda_t=\lambda_{vv}$.

\medskip

\noindent5.\quad $f=1$, $A=(u^2+1)^{-1}$:\\
\phantom{5.\quad} $\langle\p_t,\; \p_x,\; \p_v+x\p_w,\; \p_w,\; 2t\p_t+x\p_x+v\p_v+2w\p_w,\,
-v\p_x+(1+u^2)\p_u+x\p_v+\frac12(2t+x^2-v^2)\p_w\rangle$;

\medskip

\noindent6.\quad $f=1$, $A=(u^2-1)^{-1}$:\\
\phantom{6.\quad} $\langle\p_t,\,\p_{x},\,\p_{w},\,\p_v+x\p_{w},\,2t\p_t+x\p_x+v\p_{v}+2w\p_{w},\,
v\p_x+x\p_{v}-(u^2-1)\p_u+(-t+\frac{x^2}2+\frac{v^2}2)\p_{w}\rangle$;

\medskip

\noindent7.\quad $f=1$, $A=(u^2+1)^{-1}e^{\mu\arctan u}$:\\
\phantom{7.\quad} $\p_t,\; \p_x,\; \p_v+x\p_w,\; \p_w,\; 2t\p_t+x\p_x+v\p_v+2w\p_w,\,\mu t\p_t-v\p_x+(1+u^2)\p_u+x\p_v+(x^2-v^2)\p_w\rangle$;

\medskip

\noindent8.\quad $f=1$, $A=u^\mu(u+1)^{-\mu-2}$:\\
\phantom{8.\quad} $\langle\p_t,\,\p_{v},\,\p_{w},\,\p_x+v\p_{w},\,2t\p_t+x\p_x+v\p_{v}+2w\p_{w},\,
\mu t\p_t+v\p_x-u(u+1)\p_u-v\p_{v}+(\frac{v^2}2-2w)\p_{w}\rangle$;

\medskip

\noindent9.\quad $f=1$, $A=u^{-2}e^{1/u}$:\\
\phantom{9.\quad} $\langle\p_t,\,\p_{v},\,\p_{w},\,\p_x+v\p_{w},\,2t\p_t+x\p_x+v\p_{v}+2w\p_{w},\, t\p_t-v\p_x+u^2\p_u-\frac{v^2}2\p_{w}\rangle$.

Equations with parameter-functions taking the values from cases~1 and~3 are locally equivalent to the constant coefficient equations
from cases~2 and~4 correspondingly.
Reducibility of equations with parameter-functions taking the values from cases~2, 4 and 5--9 and their potential symmetries
to equations possessing Lie symmetries has been studied in~\cite{Akhatov&Gazizov&Ibragimov1987,Akhatov&Gazizov&Ibragimov1989,Ibragimov1994V1}.

This completes investigation of potential symmetries of diffusion--convection equations of form~\eqref{eqDKfh}
and consequently of form~\eqref{eqDKfgh}.

\section{Exact solutions of Fujita--Storm equation}\label{SectionOnSolutionsOfFujitaEq}

One of the possible applications of the obtained potential symmetries and potential equivalence transformations is construction of exact solutions
of equations~\eqref{eqDKfgh}.
A lot of examples of direct reductions of diffusion--convection equations with respect to potential symmetries
can be found in~\cite{Sophocleous1996,Sophocleous2000,Sophocleous2003}.
Another way of finding exact solutions for equations of form~\eqref{eqDKfgh} having nontrivial potential symmetries is
to reconstruct them from the known solutions of potentially equivalent equations
by means of application potential equivalence transformations.

Using the potential hodograph transformation~\eqref{pothodograph}
we can reduce the Fujita--Storm equation
\begin{equation}\label{eqFujStEq}
u_t=(u^{-2}u_x)_x
\end{equation}
to the linear heat equation
$\tilde v_{\tilde t}=\tilde v_{\tilde x\tilde x}$.
The same transformation connects exact solutions of these equations.

Thus, e.g., the fundamental (source) solution $v=(4\pi t)^{-1/2}e^{-x^2/(4t)}$ and
dipole solution $v=-((4\pi t)^{-1/2}e^{-x^2/(4t)})_x$
of the linear heat equation are mapped into the separable and self-similar solutions
of the Fujita--Storm equation
\begin{gather*}
u=(4\pi t)^{1/2}e^{v^2}\quad\mbox{where}\quad x=\pi^{-1/2}\int_0^ve^{-y^2}dy,\\
\mbox{and}\quad
u=x^{-1}(2t)^{1/2}\left(\ln\dfrac{1}{4\pi tx^2}\right)^{-1/2}.
\end{gather*}
correspondingly~\cite{Bluman&Kumei1980,Polyanin&Zaitsev2004}.

Applying the hodograph transformation to other known solutions
of the linear heat equation(collected, e.g., in~\cite{Olver1986}) we get the following exact solutions of the Fujita--Storm equation:
\begin{gather*}
u=c,\quad u=x^{-1},\quad u=(x-2t)^{-1/2},\quad u=(x^2\pm e^{2t})^{-1/2},
\quad u=\pm(e^{-2t}- x^2)^{-1/2},\\
u=\frac1{4\sqrt{24t^2+x}\sqrt{-6t\pm\sqrt{24t^2+x}}},\quad u=\frac t{x\sqrt{-t\ln(x\sqrt t)}},\\
u=\frac1{\sqrt{c_1^2e^{-2t}+2e^{-8t}+2e^{-4t}x}\sqrt{4-e^{8t}(-c_1e^{-t}\pm\sqrt{c_1^2e^{-2t}+2e^{-8t}+2e^{-4t}x})^2}},\\
u=\frac1{\sqrt{c_1^2e^{2t}+2e^{8t}+2e^{4t}x}\sqrt{e^{-8t}(-c_1e^{-t}\pm\sqrt{c_1^2e^{-2t}+2e^{-8t}+2e^{-4t}x})^2-4}}.
\end{gather*}

\section{Conclusion}

This work is the fourth part of the series of papers on Lie group properties of
nonlinear variable coefficient diffusion--convection equations of form~\eqref{eqDKfgh}.
In the first part of the series~\cite{Ivanova&Popovych&Sophocleous2006Part1} we construct
the equivalence transformations and perform the complete group classification of class~\eqref{eqDKfgh}.
We find not only the usual equivalence group~$G^{\sim}$ but also the extended one~$\hat G^{\sim}$
including transformations which are nonlocal with respect to arbitrary elements.
Group~$\hat G^{\sim}$ has a non-trivial subgroup of gauge equivalence transformations.
In spite of really equivalence transformations, a role of gauges in group classification
comes not to choice of representatives in equivalence classes but to choice of form of these representatives.
Application of such gauge and nonlocal transformations is important for solving
the problem of group classification in class~\eqref{eqDKfgh}. Moreover, let us note that for the class under consideration
it seems impossible to obtain complete group classification in explicit form without using transformations from
the extended equivalence group~\cite{Ivanova&Popovych&Sophocleous2006Part1,Ivanova&Sophocleous2006}.\
A similar statement is true for the problem of finding conservation laws:
classification result with respect to the usual group of equivalence transformations can be formulated in an implicit form only.
At the same time, using the extended equivalence group~$\hat G^{\sim}$,
we can present the result of classification of conservation laws in a closed and simple form.

The existence of the non-trivial subgroup of gauge equivalence transformations allows us to choose different gauges for simplification
of the solving the group classification problem. The results of group classification for class~\eqref{eqDKfgh} in two different gauges
($g=1$ and $g=h$) are adduced in the first part~\cite{Ivanova&Popovych&Sophocleous2006Part1}.\
In contrast to the most of papers on the subject under consideration we performed two essentially different group classifications:
classification with respect to the extended equivalence group and classification with respect to the set of all point transformations.
The last classification shows the way to obtain exact solutions of quite complicated equations that are reducible to simpler form
under action of point (additional) equivalence transformations. A number of examples of such solutions are adduced in the second
part~\cite{Ivanova&Popovych&Sophocleous2006Part2} of the series. We considered also in more details an interesting equation
having $sl(2,\mathbb{R})$-symmetry and constructed a set of its exact solutions by means of Lie classical method,
nonlinear separation of variables that is equivalent to a reduction with respect
to a generalized nonclassical (conditional) symmetry generator.

The next key point of the series is introducing the notion of contractions of equations, symmetries, solutions,
conservation laws and characteristics of conservation laws
(the second~\cite{Ivanova&Popovych&Sophocleous2006Part2} and the third parts~\cite{Ivanova&Popovych&Sophocleous2006Part3} of the series).
Investigation of contractions leads to deeper understanding connections between the different cases of classifications,
it can be helpful for finding exact solutions and spaces of conservation laws of contracted equations directly from the ones
of origin equations.

The notion of equivalence of conservation laws with respect to a group of transformations has been generalized in several directions:
classification of pairs ``system + space of conservation laws'';
classification of conservation laws for a given system with respect to its symmetry group;
classification of pairs ``system + a conservation law''.
Investigation of different generating sets of conservation laws of differential equations makes possible
to generalize essentially procedure of construction of potential systems, and therefore, of finding potential symmetries.
Namely, previously, for construction of simplest potential systems in cases when the dimension of the space of local conservation laws
is greater than one,  only basis conservation laws were used.
However, the basis conservation laws may be equivalent with respect to groups of symmetry transformations,
or vice versa, the number of $G^{\sim}$-independent linear combinations of conservation laws
may be greater than dimension of the space of conservation laws.
The first possibility leads to an unnecessary, often cumbersome, investigation of equivalent systems,
the second one makes possible missing a great number of inequivalent potential systems.
Class~\eqref{eqDKfgh} is very rich from this point of view. Namely, it contains nontrivial subclasses illustrating all these three possibilities.
For all these subclasses we constructed the complete lists of inequivalent potential systems and investigated their potential symmetries.

\subsection*{Acknowledgements}

NMI and ROP express their gratitude to the hospitality shown by University of Cyprus
during their visits to the University.
Research of NMI was supported by the Erwin Schr\"odinger Institute for Mathematical Physics (Vienna, Austria) in form of Junior Fellowship
and by the grant of the President of Ukraine for young scientists (project number GP/F11/0061).
Research of ROP was supported by Austrian Science Fund (FWF), Lise Meitner project M923-N13.

\end{document}